\documentclass[aps,prb,twocolumn,eqsecnum,amsmath,amssymb,floatfix]{revtex4}
\usepackage{tabularx}
\usepackage{bm}
\usepackage{euscript}
\usepackage{epsfig,psfrag,subfigure}
\usepackage{graphicx}
\usepackage{color}
\usepackage{amsfonts}
\usepackage{exscale}
\usepackage{amsbsy}
\usepackage{hyperref}

\begin{document}

\title{Theory of stripes in quasi two dimensional rare-earth tellurides}

\author{Hong Yao, John A. Robertson, Eun-Ah Kim, and Steven A. Kivelson}
\affiliation{Department of Physics, Stanford University, Stanford, CA 94305}
\date{\today}

%%%% ALIASES and NEWCOMMAND%%%%%%%%%%%%%%%%%%%%%%%%%%%%%%%%%%%%%%%%%%

\newcommand{\tperp}{t_\perp}
\newcommand{\tpara}{t_\parallel}
\newcommand{\vk}{\mathbf{k}}
\newcommand{\lang} {\langle}
\newcommand{\rang} {\rangle}
\newcommand{\tphi} {{\tilde{\phi}}}
\newcommand{\ttheta} {{\tilde{\theta}}}
\newcommand{\tPi} {{\tilde{\Pi}}}
\newcommand{\tphir} {{\tilde{\phi}}_+}
\newcommand{\tphil} {{\tilde{\phi}}_-}
\newcommand{\D}{\displaystyle}
\newcommand{\sgn}{\,\mathrm{sgn}}
\newcommand{\eA}{{\EuScript{A}}}
\newcommand{\eB}{{\EuScript{B}}}
\newcommand{\eF}{{\EuScript{F}}}
\newcommand{\eH}{{\EuScript{H}}}
\newcommand{\mH}{\mathcal{H}}
\newcommand{\vq}{\mathbf{q}}
\newcommand{\vQ}{\mathbf{Q}}
\newcommand{\Det}{\text{Det}}
\newcommand{\Tr}{\text{Tr}}
\newcommand{\td}{\textrm{d}}
\newcommand{\verts}[1]{\left\vert{#1}\right\vert}
\newcommand{\pars}[1]{\left({#1}\right)}
\newcommand{\braces}[1]{\left\{{#1}\right\}}

%%%%%%%%%%%%%%%%%%%%%%%%%%%%%%%%%%%%%%%%%%%%%%%%%%%%%%%%%%%%%%%%%%%%%%%

\begin{abstract}
Even though the rare-earth tritellurides are tetragonal materials with 
a quasi two dimensional (2D) band structure,
they have a ``hidden'' 1D character.  The resultant near-perfect  nesting of the Fermi surface  
leads to 
the formation of a charge density wave (CDW) state.  We show that for this band structure, there are two possible
ordered phases:  A bidirectional ``checkerboard''
state would occur if the CDW transition temperature were sufficiently low, whereas a unidirectional ``striped'' state, consistent with what is observed in experiment, is favored when the transition temperature is higher. This result may also give some insight into why, in more strongly correlated systems, such as the cuprates and nickelates, the observed charge ordered states are generally stripes as opposed to checkerboards.
\end{abstract}

\maketitle

\section{Introduction}\label{introduction}
Many strongly correlated (layered) materials with quasi two dimensional (2D) structure show fluctuating or static stripe order (unidirectional density wave states) over a substantial range of temperatures.\cite{Kivelson_RMP03,Tranquada04,Lilly99} 
In particular, there has been considerable interest in studying 
whether or not stripes are inextricably connected to the high temperature superconductivity that occurs in the cuprates.\cite{Tranquada04,Kivelson_RMP03} However, 
 characterizing the stripe order in these materials is difficult due to the complex and strongly correlated nature of these materials, and the presence of significant amounts of quenched disorder.\cite{Robertson06} 

From the symmetry view point, there is no difference between  a unidirectional charge density wave state (CDW) in a weakly interacting quasi-2D system and a stripe-ordered state of a strongly correlated system. CDW states can  occur in the weak coupling limit only if there are sufficiently well nested portions of the Fermi surface.  Since this is non-generic in more than 1D, one would like to identify what is special about those higher dimensional materials in which nested Fermi surfaces appear.  Moreover, for
a layered quasi-2D material with  tetragonal $(C_4)$ symmetry, the CDW ground state can either be bidirectional (checkerboards) maintaining the point group symmetry of the lattice, or unidirectional (stripes) with a reduced symmetry, and again we would like to understand what physics governs this choice.  

The rare-earth tritelluride series $R$Te$_3$ ($R$=Y, La-Tm) is particularly well suited for a detailed study of these issues.  $R$Te$_3$ consists of square Te planes alternating with weakly coupled $R$Te slabs. (See Fig. 2 of Ref.~\onlinecite{Brouet04} for example.)
 The electronically active valence band derives from the 5$p$ orbitals of the planar Te atoms, and is thus expected to be relatively weakly correlated.  The existence of a unidirectional CDW was first detected by transmission electron microscopy (TEM).\cite{DiMasi95,DiMasi96}
 Angle resolved photoemission spectroscopy (ARPES)~\cite{Gweon98,Brouet04,Komoda04} measurements have shown that the CDW ordering wavevector nests large portions of the Fermi
 surface, which are in turn gapped, thus strongly indicating that the CDW is a consequence of  Fermi surface nesting.  
More recent X-ray scattering and STM measurements have confirmed the existence of the CDW and its unidirectional character in great detail.\cite{Kim06, Fang06}
Detailed studies of the Fermi surface topology using ARPES~\cite{Brouet04} and positron annihilation~\cite{Laverock05} support the notion that the CDW is the result of a nesting-driven Fermi surface instability.   However, the driving force for the strong breaking of the point group symmetry ($C_4 \rightarrow C_2$) produced by the unidirectional CDW formation in $R$Te$_3$ has not been clear previously. 

We shall show in the present paper that the nesting of the Fermi surface of $R$Te$_3$ reflects a ``hidden'' 1D character of the electronic band structure which derives from the highly anisotropic hopping amplitude of the Te $p_x$ and $p_y$ orbitals.\cite{footnote1}
Consequently, CDW order occurs for a dimensionless effective interaction, $\lambda$, in excess of an extremely small critical value, $\lambda_c \ll 1$.
(In 1D, the fact that $\lambda_c=0$ is the famous Peierls instability.)
Moreover, we shall show that there are two possible patterns of CDW order that can take best advantage of the nesting:
checkerboard order that occurs for $\lambda$ slightly larger than $\lambda_c$, and stripe order, which is also rotated relative to the checkerboard order, which occurs for $\lambda> \lambda_0\approx \lambda_c[1 + {\cal O}(\lambda_c)]$; in more physical terms, what this means is that when the CDW transition temperature, $T_c$, or the CDW gap, $\Delta_0$, is sufficiently large, stripe order is favored.
(The resulting phase diagram is shown in Fig. 4.)

\section{Band structure considerations}\label{SEC-bandstructure}
It is well established from transport measurements\cite{Dimasi94,Ru06} that the coupling between layers in $R$Te$_3$ is small. Comparisons between different rare earth compounds confirm the minor role of the rare earth in the electronic structure.\cite{Brouet04,Ru06} 
Hence the physical properties of $R$Te$_3$ are dominantly determined by
the Te planes common to all $R$Te$_3$. It is therefore natural to 
consider a simple model of the electronic structure 
of a single Te plane.

\begin{figure}[t]
\includegraphics[scale=0.25]{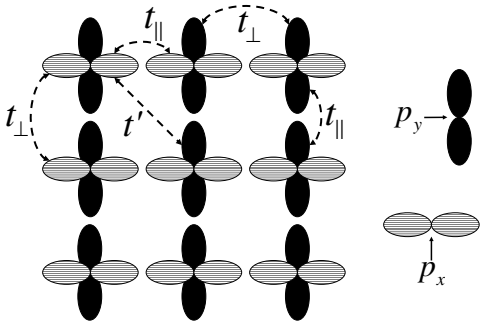}
\caption{The $p_x$ (hatched) and $p_y$ (solid) orbitals in the single Te square lattice. 
Due to the highly anisotropic profile of the $p$ orbitals, the hopping amplitude, $\tpara$, along the extended direction of the given $p$ orbital is much larger than the hopping amplitude $\tperp$ along the direction perpendicular to the extended direction. $t^\prime$ is a second-neighbor hopping, which is the shortest-range interaction which mixes the $p_x$ and $p_y$ bands, and which we neglect for present purposes.}
\label{telluridelattice}
\end{figure}

It is known, 
from first principle band structure calculations,\cite{Kikuchi98} 
 that the energy of the $5p_z$ is shifted by crystal field effects so that it lies well below the Fermi energy. Consequently,  
the relevant bands close to the Fermi surface are the  $5p_x$ and $5p_y$ orbitals, which are approximately 5/8 filled.  The crystallographic unit cell contains two Te atoms per plane due to a
 slight inequivalence between two sublattices produced by the $R$Te slab orientation and the crystal structure is very weakly orthorhombic;\cite{Laverock05} for simplicity, we will ignore these subtleties and consider an idealized Te square lattice. 
The band structure can be well approximated in the tight-binding approximation;  the resulting Hamiltonian is shown schematically in Fig.~\ref{telluridelattice}. 

Following Ref.~\onlinecite{DiMasi95}, we 
neglect all interactions of longer-range than the nearest neighbor hopping, with the consequence that  there is no hybridization between the $p_x$ and $p_y$ bands.  (Small further neighbor interactions, such as the second neighbor hopping $t^\prime$ in Fig.~\ref{telluridelattice}, have a large effect on the Fermi surface topology
where the two bands intersect, but can readily be shown to have little effect on the results obtained below.) 
Working in units where the lattice constant of the square lattice $a\!=\!1$, the dispersion for the $p_x$ band and $p_y$ band can be readily derived: 
\begin{equation}
\begin{split}
\epsilon_{\vk,p_x}&=-2\tpara \cos(k_x)+2\tperp \cos(k_y)\\
\epsilon_{\vk,p_y}&=\phantom{-}2\tperp \cos(k_x)-2\tpara \cos(k_y),
\end{split}
\end{equation}
where  $\tpara$ ($\tperp$)  represents the hopping amplitude parallel (perpendicular) to the extended direction of the given $p$ orbital. Due to the highly anisotropic profile of the $p$ orbital electron wave function,  
the hopping amplitude $\tpara$, along the extended direction of the given $p$ orbital, is much larger than the hopping amplitude $\tperp$ perpendicular to the extended direction.
Indeed  these hopping amplitudes have been estimated~\cite{Kikuchi98} to be $\tpara\approx 2.0{\textrm eV}$, $\tperp\approx 0.37{\textrm eV}$,  and $t^\prime\approx 0.16$eV
 for $R$Te$_3$.  Thus, it is reasonable to set $t^\prime=0$ and to treat $\tperp/\tpara \approx 0.18$ as a small parameter.

\begin{figure}[t]
\subfigure[]  {
\includegraphics[width=0.217\textwidth]{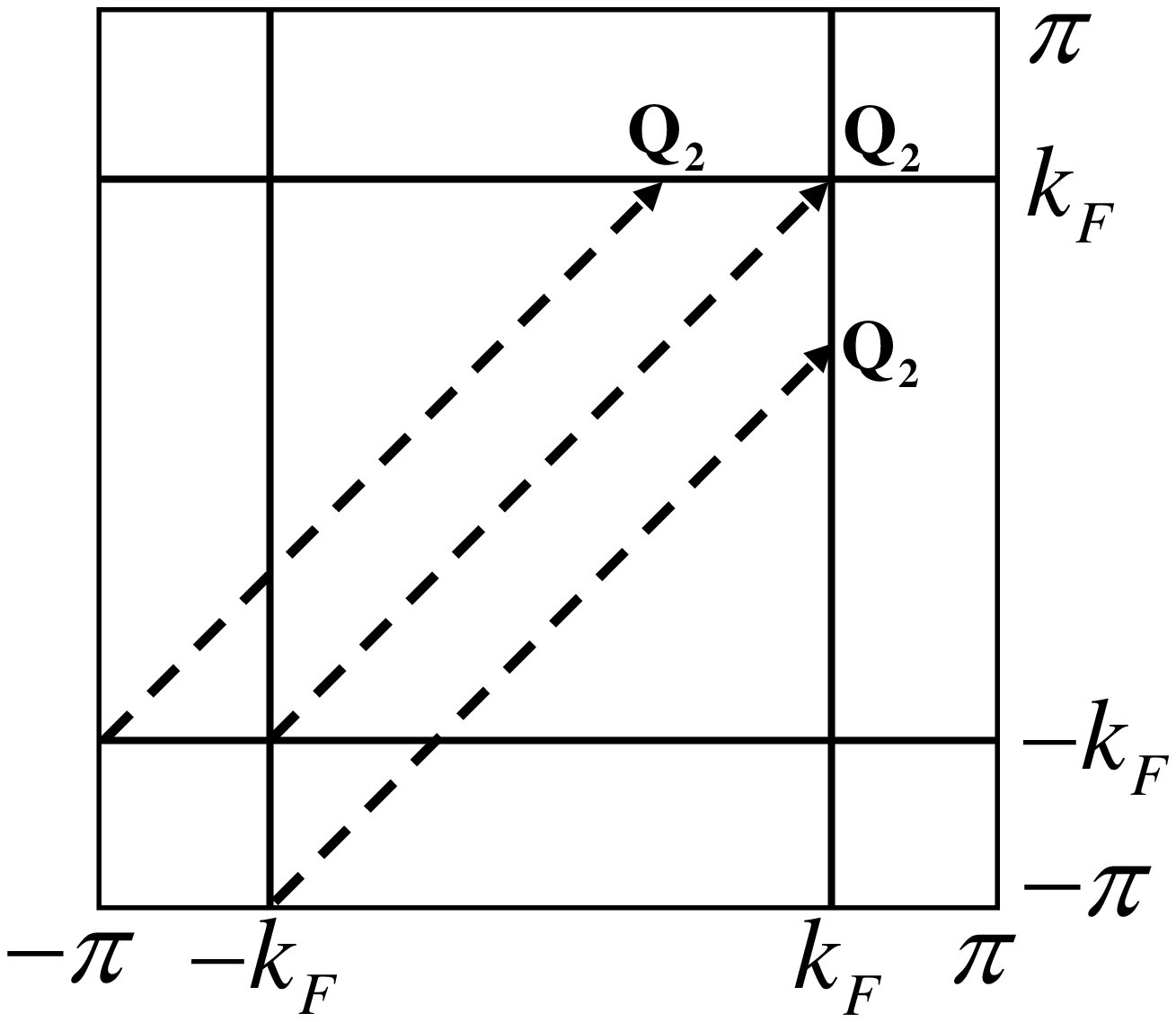} 
}
\subfigure[]{
  \includegraphics[width=0.22\textwidth]{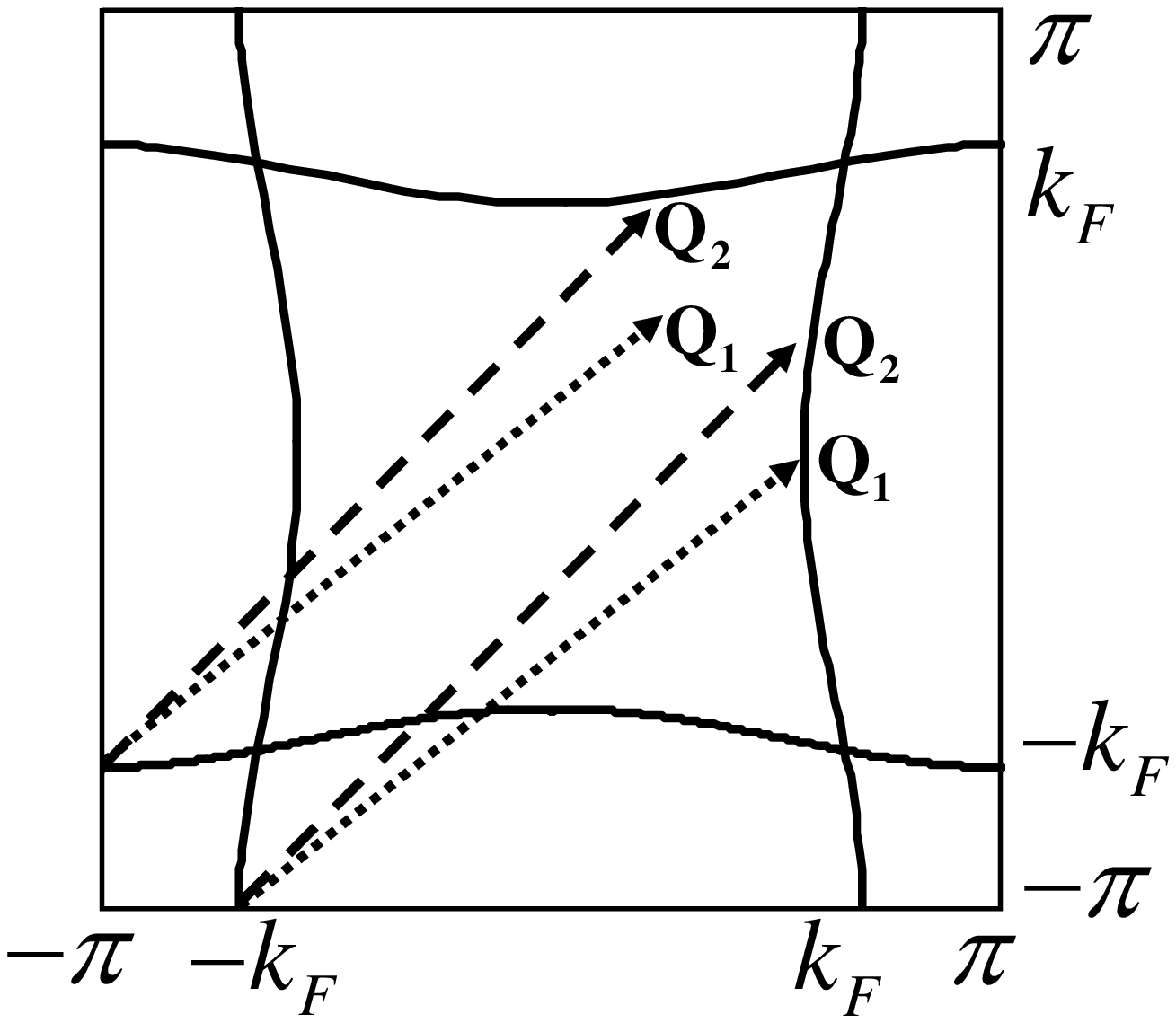} 
}
\caption{Fermi surfaces (a) for $\tperp=0$ and (b) for $\tperp$ small, but non-zero.
For $\tperp=0$, the wavevector ${\bf Q_2}\equiv(2k_F,2k_F)$ nests both the  $p_x$-FS and the $p_y$-FS perfectly. 
For non-zero $\tperp$, the wavevector ${\bf Q}_1\equiv (2k_F,\pi)$ does a much better job of nesting the $p_x$-FS  but a poor job  on the $p_y$-FS. On the other hand, ${\bf Q}_2$ does a moderately good job of nesting the entire FS.}
\label{FIG-FermiSurface}
\end{figure}

The small magnitude of the ratio $\tperp/\tpara$
implies a secret quasi-1D character of the band structure. 
For $\tperp\!=\!0$, the 
system would be equivalent to an array of 1D wires.
Nonetheless, the system would maintain overall $C_4$ symmetry since these 
%the array of 
$p_x$ and $p_y$ ``wires'' are perpendicular to each other.
Thus, even in this limit,
the system would not display 2D anisotropy in any transport measurement. However, 
the resulting band structure would consist of 
two parallel 1D FS's as shown in Fig.~\ref{FIG-FermiSurface}(a). With $k_F$ defined by the implicit relation $\mu=-2\tpara\cos{k_F}$, where $\mu$ is the chemical potential, any wavevector $(\pm 2k_F,k_y)$ would then perfectly nest the $p_x$-FS for arbitrary $k_y$ and wavevectors $(k_x,\pm 2k_F)$ would perfectly nest the $p_y$-FS for any $k_x$.
In particular, the wavevectors $(\pm 2k_F, \pm 2k_F)$ perfectly nest both the $p_x$ and $p_y$ FS's.

A small, but nonzero $\tperp$ introduces nonzero but small curvature to the FS (see Fig.~\ref{FIG-FermiSurface}(b));
the $p_x$-FS is determined by 
\begin{eqnarray}
k_x=\pm\left[k_F -\left(\frac{\tperp}{\tpara}\right)\frac{\cos k_y}{\sin{k_F}}+ {\cal O}\left({\tperp^2\over \tpara^2}\right)\right],
\end{eqnarray}
where $k_F$ is defined above.
Here $+(-)$ represents the right (left) portion of the $p_x$-FS respectively. 
Obviously, the $p_y$-FS can be obtained from the $p_x$-FS by $90^\circ$ rotation.

As a consequence of the finite curvature, the wavevector $(\pm 2k_F,k_y)$  nearly perfectly nests the $p_x$-FS {\it only for} $k_y\!=\!\pi$, 
and equivalently the $p_y$-FS is best nested by $(\pi,\pm 2k_F)$. However, away from the half filling, 
$(\pm 2k_F,\pi)$ does a poor job in nesting the $p_y$-FS and 
conversely for $(\pi,\pm 2k_F)$.
Note that $(2k_F,\pi)$, $(-2k_F,\pi)$, $(\pi,2k_F)$ and $(\pi,-2k_F)$ are related by the $C_4$ symmetry of the host lattice.
On the other hand, 
the wavevector $(2k_F,2k_F)$ (and its $C_4$ symmetry related vectors), which perfectly nests the full FS for $\tperp\!=\!0$, 
does a reasonable job in nesting both the $p_x$-FS and the $p_y$-FS.
Thus, for small $\tperp/\tpara$, there are two independent candidates (not related to each other via $C_4$ symmetry) for the CDW ordering wavevectors: 
\begin{equation}
{\bf Q}_1\equiv (2k_F,\pi) \ \ \ \ \ \ \ {\rm and}\ \ \ \ \ {\bf Q}_2\equiv (2k_F,2k_F).
\end{equation}

To see which ordering vector is preferred, we now compute the corresponding charge density (Lindhard) susceptibility, $\chi(\vq ;T)$.
For a given wavevector 
$\vq$ at a given temperature $T$, 
\begin{equation}
\chi(\vq ;T)=-2\sum_{\eta}\int\frac{d\vk}{(2\pi)^2}
\frac{f(\xi_{\vk\eta},T)-f(\xi_{\vk+\vq\eta},T)}
{\xi_{\vk\,\eta}-\xi_{\vk+\vq\,\eta}},
\label{eq:Lindhard}
\end{equation} 
where the integral is over the first Brillouin zone, $\eta\!=\! p_x, p_y$ is the band index, and $\xi_{\vk\eta}=\epsilon_{\vk\eta}-\mu$. The factor of 2 is from two spin polarizations of the electrons and $f(\xi_{\vk\eta},T)$ denotes the  Fermi-Dirac distribution function. 

For small $\tperp/\tpara$, 
the susceptibilities at these two vectors have the following approximate analytical forms, which are valid for temperatures $T
\ll \tperp$ (in units $k_B=1$):
\begin{eqnarray}
&&\chi({\bf Q}_1;T)\approx\rho(E_F) \log\left[{\frac{\alpha_0\tpara}{T+\alpha_1 \tperp^2/\tpara}} \right] \nonumber \\
&&~~~~~~~~~~~~~~~~~~~+\rho(E_F)\log\left[{\frac{\alpha_0\tpara}{T+\alpha_2 \tpara}}\right], \label{eq:chi1}\\
&&\chi({\bf Q}_2;T)\approx 2\rho(E_F) \log\left[{\frac{\alpha_0\tpara}{T+\alpha_3 \tperp}}\right],\label{eq:chi2}
\end{eqnarray} 
where $\rho(E_F)\equiv [{2\pi \tpara \sin{k_F}}]^{-1}$ is the density of states per spin per band and
$\alpha_0\tpara$ is the ultraviolet cutoff, which is of order of the bandwidth ($\alpha_0\approx 2$). The parameters $\alpha_i$ are defined by
\begin{eqnarray}
&&\alpha_1\approx |\cos{k_F}|/(4\sin^2{k_F}), \nonumber\\
&&\alpha_2\approx 2|\cos{k_F}|,  \\
&&\alpha_3\approx |\cos{k_F}|.\nonumber
\end{eqnarray}
%Since i
In $R$Te$_3$, the bands are approximately  5/8 filled, thus $k_F\approx 5\pi/8$.
In the small but non-zero $\tperp/\tpara$ limit, 
\begin{equation}
\chi({\bf Q}_1;0)\approx \ \chi({\bf Q}_2;0) + \rho(E_F)\log[2\sin^2(k_F)] .
\end{equation}
Therefore, if the  CDW transition occurs at low enough temperatures  the wavevector ${\bf Q}_1$ will be favored so long as $\pi/4 \lesssim k_F \lesssim 3\pi/4$. 
However, at high enough temperatures, $T> T_0\sim \tperp^2/\tpara$,
$\chi({\bf Q}_2;T)$ is greater than  $\chi({\bf Q}_1;T)$. 
Therefore, we expect there to be a finite temperature $T_0$ such that if the CDW transition temperature $T_c\!<\!T_0$, the ordering vector is ${\bf q}\approx{\bf Q}_1$, while for $T_c\!>\!T_0$,  ${\bf q} \approx {\bf Q}_2$.

\begin{figure}[t] 
\includegraphics[width=0.4\textwidth]{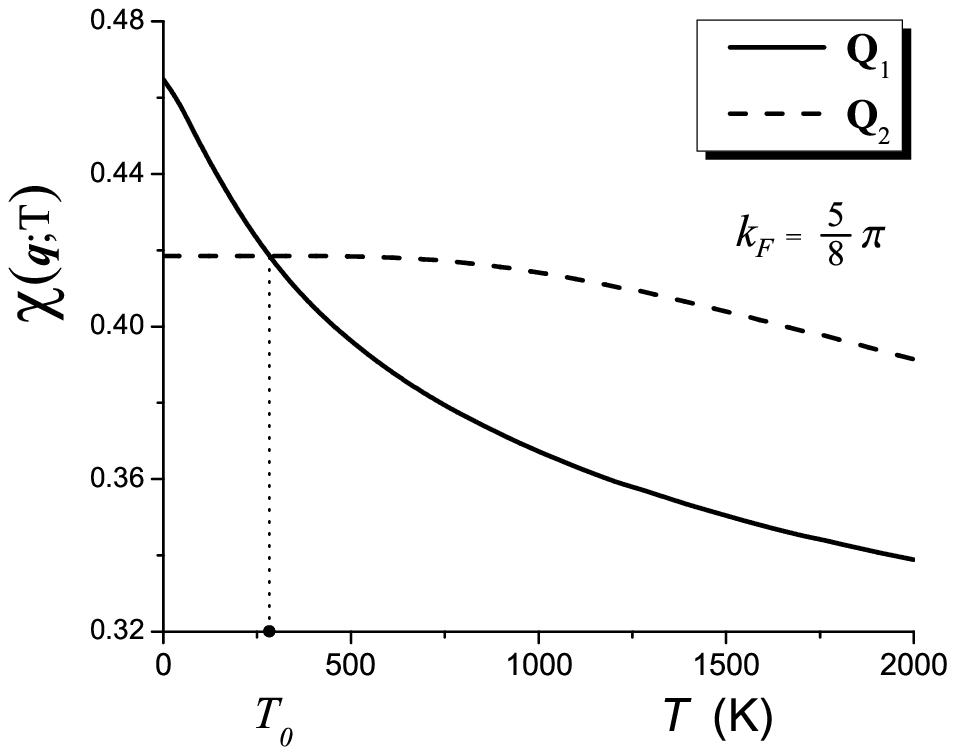}
\caption{The Linhard susceptibility $\chi(\vq;T)$ as a function of $T$ for $\vq={\bf Q}_1\equiv (2k_F,\pi)$ and
$\vq={\bf Q}_2\equiv (2k_F,2k_F)$.}
\label{FIG-SusceptibilityCompare}
\end{figure}

For a more quantitative analysis, we have numerically evaluated the susceptibility $\chi(\vq;T)$ of Eq.~\eqref{eq:Lindhard}, for wavevectors $\vq$ over the whole Brillouin zone.
The calculated susceptibility  has a maximum at a wavevector $\vq_{max}$ close to ${\bf Q}_1$ or 
${\bf Q}_2$, depending on whether the temperature, $T<T_0$ or $T>T_0$, in agreement with  our analysis of the FS. 
The precise location of $\vq_{max}$ shifts slightly as a function of $T$, but so long as $T \ll \tpara$, the maximum always lies close to  
${\bf Q}_j$, so this deviation does not significantly affect the 
 resulting picture of the phase diagram.  We will henceforth neglect this small effect, and  focus our discussion on the wavevectors ${\bf Q}_j$. Fig.~\ref{FIG-SusceptibilityCompare} shows the calculated $\chi(\vq;T)$ for $\vq\!=\!{\bf Q}_j$.

This result can be understood as follows:
At low temperatures, the states arbitrarily close to the nearly perfectly nested portions of the Fermi surface dominate the susceptibility favoring ${\bf Q}_1$.  However, for $T > T_0$,
the curvature of the FS has negligible effect, so ${\bf Q}_2$ 
is preferred.  
Since ${\bf Q}_1$ nests only the $p_x$ portion of the FS, there is little competition between the ordering tendency at ${\bf Q}_1$ and at the symmetry related vector, ${\bf \bar Q}_1=(\pi,2k_F)$.  Thus, as we shall show below, wherever CDW ordering with wavevector ${\bf Q}_1$ occurs, simultaneous ordering occurs at ${\bf \bar Q}_1=(\pi,2k_F)$, making this a state with ``checkerboard'' order.  However, since ordering at wavevector ${\bf Q}_2$ opens a gap on the entire FS, it prevents ordering at the symmetry related vector, ${\bf \bar Q}_2=(-2k_F,2k_F)$;  when ${\bf Q}_2$ is the preferred ordering vector, the resulting state has ``stripe'' order.

\section{Mean-Field Phase Diagram} \label{SEC-Model}

To obtain an explicit phase diagram, we add an electron-phonon coupling and solve the resulting model in mean-field approximation.
Thus, we 
investigate  
 the following mean-field Hamiltonian
\begin{eqnarray}\label{MFH}
H_{MF}=\sum_{\vk,\eta}\xi_{\vk\eta}c^{\dag}_{\vk\eta}c_{\vk\eta}
+\sum_{\vq}\frac{\rho(E_F)}{2\lambda_{\vq}} |\Delta_{\vq}|^2 \qquad \nonumber\\
+\sum_{\vk,\vq,\eta}\Delta_{\vq} c^{\dag}_{\vk+\vq\eta}c_{\vk\eta},
\end{eqnarray}
where the sum over $\vq$ runs over the possible ordering vectors, $\vq = \pm {\bf Q}_1,\  \pm {\bf \bar Q}_1,\  \pm {\bf Q}_2,\ {\rm and}
 \pm {\bf \bar Q}_2,$ and all harmonics, such as $\pm 2{\bf Q}_j$ and $\pm {\bf Q}_j\pm {\bf \bar Q}_j$. The order parameter $\Delta_{\vq}$ for the CDW with the ordering vector $\vq$
is related to the lattice distortion $\langle x_\vq \rangle$ through
$\Delta_{\vq}=\alpha_{\vq}\langle x_\vq \rangle$,
where $\alpha_{\vq}$ is the electron-phonon coupling, 
and the dimensionless effective interaction $\lambda_{\vq}=[\alpha_{\vq}^2\rho(E_F)]/[M\omega^2_{\vq}]$, where $\omega_\vq$ is the phonon frequency and $M$ is the ion mass.  The self-consistent value of the gap is obtained by minimizing the free energy computed from $H_{MF}$ with respect to $\Delta_\vq$. 
Note that $\Delta_{-\vq}=\Delta_{\vq}^{\ast}$. We use the approximation $\lambda_\vq=\lambda$ in the rest of this paper.  

At high temperatures, clearly $\Delta_{\vq}=0$ for all $\vq$.  
As the temperature is lowered, for sufficiently large $\lambda$, solutions with non-zero values of $\Delta_\vq$ appear.  In general, when $\Delta_\vq$ is non-zero at some wavevector, $\vq$, it is non-zero, although possibly much smaller, at all harmonics, as well.

For commensurate order, the mean-field equations can, in principle, be evaluated numerically exactly, but for high order commensurability this is quite a complicated problem, as there are many harmonics, and hence many possible patterns of order.  For incommensurate order, even this is not possible.  

Fortunately, in the present case, for the most part the transitions are continuous, or at worst weakly first order.  Therefore, the phase boundaries can be identified by expanding the mean-field (Landau) free energy to low order in powers of $\Delta_\vq$.  The coefficients in this expansion are expressed as convolutions of free Green functions, as derived explicitly up to fourth order in the Appendix. To this order, only the fundamentals and their second harmonics enter the expansion;  if $\vq$ and $\bar\vq$ are two (symmetry related) fundamental ordering vectors, then the higher harmonics in powers of the fundamental, satisfy $\Delta_{n\vq+m\bar\vq}\sim [\Delta_{\vq}^n\Delta_{\bar\vq}^m]$.  Indeed, in this context, we can treat the harmonics as slaves to the fundamentals and obtain a Landau free energy expressed exclusively in terms of $\Delta_\vq$ and $\Delta_{\bar\vq}$, as is done explicitly in the Appendix.  The result is a free energy density
\begin{eqnarray}
&&\mathcal{F}=\sum_{j=1}^{2}  \Big[ r_j(|\Delta_{\vQ_j}|^2+|\Delta_{\bar \vQ_j}|^2) \\  &&~~+\frac{u_j}{4}(|\Delta_{\vQ_j}|^2+|\Delta_{\bar\vQ_j}|^2)^2
+\gamma_j|\Delta_{\vQ_j}|^2|\Delta_{\bar\vQ_j}|^2 \Big] +\ldots,\nonumber
\end{eqnarray}
where 
\begin{equation}
r_j=\rho(E_F)/\lambda-\chi(\vQ_j;T), 
\end{equation}
and 
$\ldots$ signifies both higher order terms in powers of $\Delta$, as well as unimportant biquadratic terms, such as $|\Delta_{\vQ_1}|^2|\Delta_{\vQ_2}|^2$, which come into play only if ordering occurs both at $\vQ_1=(2k_F,\pi)$ and $\vQ_2=(2k_F,2k_F)$.
The coefficients $u_j$ and $\gamma_j$ are evaluated in the Appendix.  

The mean-field phase diagram in Fig.~\ref{FIG-phasediagram} was constructed from this expression as follows:  The phase boundaries were determined as the points at which either $r_1$ or $r_2$ vanishes, so both $r_1$ and $r_2 > 0$ in the high temperature phase.  Throughout the ordered region of the phase diagram, we always find that $\gamma_1 < 0$ and $\gamma_2 >0$, so wherever $\vQ_1$ ordering is favored, simultaneous ordering at $\vQ_1$ and ${\bar \vQ_1}$, {\it i.e.}, checkerboard order, occurs, while in the regions where $\vQ_2$ ordering is favored, either $\Delta_{\vQ_2}$ or $\Delta_{\bar\vQ_2}$ is non-zero, but not both, {\it i.e.}, striped order occurs.  The transition to the stripe phase is continuous so long as $u_2 >0$, which is the case at large coupling.  The tricritical point occurs where $u_2$ vanishes.  Strictly speaking, where the transition becomes first order (the dashed lines in the figure), the Landau expansion is inadequate;  the dashed lines should therefore be interpreted as schematic representations of the true phase boundaries.  In evaluating the coefficients in the Landau expansion, 
we use the band structure with the representative values given in the caption of Fig.~\ref{FIG-phasediagram}, but the phase diagram is qualitatively similar for any $\tperp \ll \tpara$ and $\pi/4 \lesssim k_F \lesssim 3\pi/4$.

\begin{figure}[t]
\includegraphics[scale=0.35]{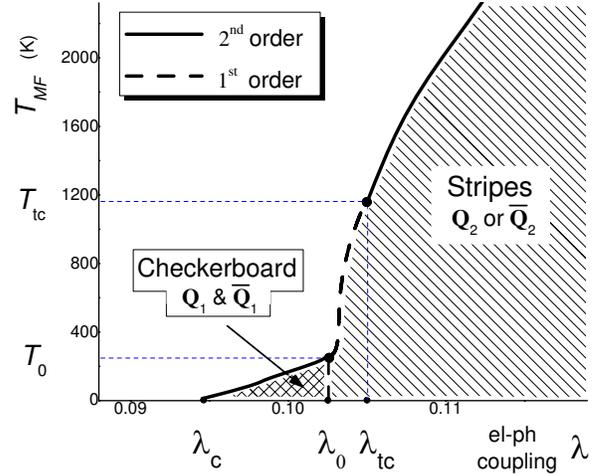}
\caption{The mean-field phase diagram.
For $\lambda\!>\!\lambda_0$, the system orders to a stripe phase below the transition temperature. The associated phase transition is  second
order for  $\lambda\!>\!\lambda_{\textrm{tc}}$ and  first order for $\lambda_0\!<\!\lambda\!<\!\lambda_{\textrm{tc}}$.
For $\lambda_c\!<\!\lambda\!<\!\lambda_0$, the system orders to a  checkerboard pattern.  For $\lambda<\lambda_c$, there is no CDW instability down to
zero temperature. Explicit numbers refer to the representative values of $\tpara=2.0$eV, $\tperp=0.37$eV and $k_F\approx 5\pi/8$. $T_{MF}$ signifies the mean field $T_c$; the actual $T_c$ will be smaller due to fluctuation effects.}
\label{FIG-phasediagram}
\end{figure}

The main new physics apparent in this phase diagram arises from the sign of $\gamma_j$.  The fact that $\gamma_2$
 is robustly positive follows from the fact that $\vQ_2$ does a moderately good job of nesting the entire FS;  in the presence of a non-zero $\Delta_{\vQ_2}$, most of the FS is gapped, so there are no remaining portions of the FS to be nested by ${\bar{\bf \vQ}_2}$.  Conversely, the fact that $\vQ_1$ fails so badly to nest the $p_y$ portions of the FS, implies that there is no substantial interference between $\vQ_1$ and ${\bar \vQ_1}$ ordering;  this leads to the expectation of a near vanishing biquadratic coupling, or in other words, $\gamma_1\approx -u_1/2< 0$, an expectation that we find is approximately satisfied.  (At $T=0$, $\gamma_1\approx -25$ and $u_1\approx 47$.)

Using the Landau free energy with coefficients given in the Appendix, for the band structure with the representative values, we can estimate the various critical values of the of the coupling constant:  the minimum value for the existence of a CDW ground state is $\lambda_c \approx 0.093$;  the critical value that separates the regime of checkerboard order from stripe order is $\lambda_0\approx 0.103$;  the tricritical coupling is $\lambda_{tc}\approx 0.105$; the temperature at the tricritical point is $T_{tc}\approx 0.098$\,eV. The maximum checkerboard ordering temperature, $T_0$, which occurs on the edge of the first order transition to the striped phase, is the one qualitative feature of the phase diagram that depends on the higher order terms in the Landau expansion that we have not computed.  The estimate for $T_0$ given in the figure comes from the assumption that the first order phase boundary lies close to the point at which $r_2$ vanishes.

\section{fluctuation effects}
Because of the hidden 1D character of the ordering we have been exploring, we expect fluctuation effects will have a large quantitative effect on the mean-field phase diagram.  Indeed,
it is
common in quasi 1D CDW systems for the ordering temperature, $T_c$ to be significantly suppressed below its mean-field value.  This is reflected in larger values (than the  prediction of mean-field theory) of the ratio,  $\Delta_0/T_c$, of the zero temperature gap to
the actual $T_c$.\cite{Johnston85,Kwok90}

We can estimate the extent to which fluctuations suppress $T_c$ in two different limits: If the anisotropy is not too large, {\it i.e.} if $\tpara\gg\tperp\gg\Delta_0$, the suppression is fractionally small, and can be estimated, as in the theory of superconductivity, by the Brout criterion, $T_{MF} -T_c\sim T_{MF}\Delta_0^2/\tpara\tperp$.  Conversely, if $\tpara\gg\Delta_0\gg \tperp$, while $\Delta_0$ may still be crudely determined by mean-field considerations, $T_c \ll T_{MF}$ and is determined by entirely different physics.  In this limit, the intra-chain CDW fluctuations can be treated using the theory of the one dimensional electron gas, and $\tperp$ can be included in the context of inter-chain mean-field theory, with the result that $T_c \sim \tperp(\tperp/\tpara)^\alpha$ where $\alpha$ is an interaction dependent constant.\cite{Scalapino75,Lee73} Whenever there is a large fluctuational suppression of $T_c$, local CDW correlations are expected to survive in a broad range of
temperatures above $T_c$;  roughly, the local CDW correlations develops in the temperature range $T_c < T < T_{MF}$, where $T_{MF}$ is the mean-field transition temperature. 
 
For $R$Te$_3$, $\Delta_0\sim 260-400$meV,\cite{Brouet04} corresponding to a mean transition temperature $T_{MF}\sim 1500-2000$K and $\lambda\sim 0.1$. 
Since $\tperp \approx 0.38$eV $ \sim \Delta_0$, a reliable estimate of $T_c$ is not possible for these materials, although a large supression relative to the mean-field value can be expected. 
An expression due to Barisic\cite{Barisic} based on physically plausible but hard to justify approximations yields $T_c\sim (\tperp/\tpara)T_{MF}\approx 300$K. Indeed, very recently, it has been found\cite{Ru06_unpublished} that he CDW phase transition in $R$Te$_3$ occurs (depending on $R$) in the range $T_c=260-400$/,K, and substantial CDW correlations persist well above $T_c$.

One consequence of remaining local CDW correlations above $T_c$ will be the presence of peaks in the structure factor at
positions corresponding to the Bragg vectors of the ordered state,
but with width inversely proportional to the thermal correlation
length.\cite{Girault89} However in the case of a stripe phase, more dramatic effects can be expected. If there is only a single
transition; then the ``stripe liquid" phase above $T_c$ does not
break any of the lattice symmetries, and consequently equivalent
peaks in the structure factor should appear both at the stripe
ordering vector, and at the conjugate wavevector (rotated by 90$^\circ
$).  Such behavior has already been seen, albeit only in the magnetic
scattering, in the stripe liquid phase of La$_{2-x}$Sr$_{x}$NiO$_4$.\cite{Lee_Tranquada02} On the other hand, a two stage transition is also possible with an intermediate, ``stripe nematic" phase,\cite{Kivelson_Fradkin_Emery_Nature} in which stripe fluctuations are sufficiently violent to
restore translation, but not the full $C_4$ symmetry of the host
crystal; in such a phase, the peaks at the stripe ordering wave
vector should be stronger (possibly, much stronger) than those at the
conjugate wavevector. In this case, there must be a second
transition at still higher temperatures to an isotropic state.

If future experiments can confirm the predictions above regarding the fluctuation effect, 
they will provide an important laboratory for exploring
the physics of a stripe liquid, with possible broader implications to
many stripy materials.

\section{concluding remarks}\label{SEC-Conclusion}

The present mean-field theory gives results that are broadly consistent with experimental observations in $R$Te$_3$. From the value of
the CDW gap measured by ARPES,\cite{Gweon98,Brouet04} one can estimate 
the mean-field transition temperature to be 
$T_{MF} \approx 2000$\,K, which is an order of magnitude greater than the theoretical value of $T_0$ and also greater than  $T_{tc}$.  Hence, the theory predicts that   $R$Te$_3$ should have a unidirectional CDW  ordered phase (stripes) with ordering vector approximately equal to either $\vQ_2=(2k_F,2k_F)$ or ${\bar \vQ_2}=(2k_F,-2k_F)$.\cite{footnote2} 
This is consistent with the experiments\cite{DiMasi95,Gweon98,Brouet04} which find stripe order with $\vQ_\textrm{CDW}\approx (0.71\pi,0.71\pi)$ in the present non-doubled unit cell notation.\cite{footnote3} With the simplified band-structure used in the present analysis,  $2k_F\approx 3\pi/4$ (mod $2\pi$), However, as mentioned above, the maximum of the susceptibility occurs at a slightly different weakly temperature dependent wave-vector;  at $T= 350$K, the maximum occurs at $\vq_{max}=(0.72\pi,0.72\pi)$, which is almost identical to the experimental results at comparable temperatures. Moreover, the mean field theory predicts that the phase transition into the stripe ordered phase from higher temperature phase is a second order phase transition. This is also consistent with recent experimental results.\cite{Ru06_unpublished}

Even though we are dealing with a stoichiometric phase with an integer number of electrons per unit cell, if the transition is continuous, then (at least for commensurability $N >4$) the CDW ordered state just below $T_c$ is generically incommensurate, in the sense that the ordering vector is not locked to the lattice.  This follows from the fact that the lowest order term which locks the CDW to the lattice is proportional to $\Delta^N\cos(N\phi)$  and so is irrelevant (for $N >4 $) in the renormalization group sense\cite{Coppersmith82} at $T_c$.   (Here $\Delta$ is the order parameter and $\phi = 2k_Fx_0$ determines the relative phase between the CDW and the underlying crystalline lattice.)  In the present problem, two effects contribute to shift the ordering vector from its commensurate value.  Firstly, even at $T=0$, there is a shift of $k_F=3\pi/4[ 1 + {\cal O}(\tperp/\tpara)^2]$ (which we have not discussed explicitly) due to the 2D dispersion. Moreover, $\vQ_2$ depends weakly on the temperature and (as mentioned above) is noticeably different from its $T=0$ value at $T=T_c$. However, it is possible that, under appropriate circumstances, there will be a second commensurability lock-in transition with critical temperature $T_{Com} < T_c$.

Another subtlety of the problem derives from the existence of Te bilayers in $R$Te$_3$.  If the bilayer splitting, $t_{bil} \ll \Delta_0$, it can be ignored, and a single transition occurs as treated in the present work.  However, if $\tpara \gg t_{bil} \gg \Delta_0$, we should treat the ordering in the bilayer split bands separately.  In this case, there should be two distinct ordering transitions, at distinct ordering temperatures, with slightly different ordering vectors, $k_{F,\pm}=k_F\pm t_{bil}/v_F$.  

Recent results\cite{Ru06_unpublished} on RTe$_3$ with different rare earth elements, R, show that in some cases there does, in fact, appear to be a second phase transition at temperatures below $T_c$.  For various reasons, we conjecture that where the second transition occurs, it is due to the bilayer splitting, but whether it is this, or a commensurate lock-in, or some other transition remains unsettled.

Charge density waves often break the point group symmetry of the host lattice. This phenomenon is particularly common in strongly correlated materials including many transition metal oxides which exhibit stripe order.  The strong interactions present in those ($d$-band) systems make the physics of pattern selection more difficult to study from a microscopic viewpoint.  However, one may hope that extrapolating from the weak coupling limit (solved in the present paper) may give some insight into the deeper issues of pattern selection in highly correlated materials. 

The underlying physics that is responsible for the existence of a CDW state in the present class of reasonably weakly interacting quasi-2D systems is the existence of a hidden 1D character of the band structure.  In the present paper, we have explored only the grossest aspects of this structure - primarily those amenable to a mean-field analysis.  At temperatures or frequencies in excess of the ordering temperature, where the band structure can be approximated as that of intersecting 1D systems, it is probable that more interesting fluctuation effects, associated with the breakdown of Fermi liquid theory in 1D, should be observable.  Indeed, some evidence already exists\cite{Giamarchi_Sacchetti} from high energy spectroscopies of anomalous power-law behaviors reminiscent of the 1D Luttinger liquid.  We are currently exploring this aspect of the problem.
\newline
\newline
\noindent{\bf Acknowledgments} We would like to thank S. Brown, A. Fang, I. R. Fisher, A. Kapitulnik, N. Ru and K. Y. Shin, for helpful discussions, and J. Allen and G.-H. Gweon for partially inspiring this work. This work was supported in part by NSF grant number DMR0531196 at Stanford University.
\vspace{-0.2cm}
\appendix
\section{The linked cluster expansion}
\begin{figure}[th]
\includegraphics[scale=0.4]{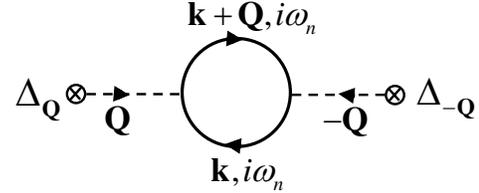}
\caption{The second order Feynman diagram from which we obtain the Lindhard susceptibility, $\chi_{\vQ}$.
Higher order diagrams differ only in the number of external legs.}
\label{FIG-chi}
\end{figure}
Here we sketch the details of the linked cluster expansion that was used to obtain
the coefficients of the GL free energy. We treat the off-diagonal part of the mean
field Hamiltonian Eq.~\eqref{MFH} 
\begin{equation}
V = \sum_{\vk,\vq,\eta}\Delta_{\vq} c^{\dag}_{\vk+\vq\eta}c_{\vk\eta}
\end{equation}
as a perturbation to obtain the Landau free energy:
\begin{eqnarray}\label{Eq-GLclusterexpansion}
\mathcal{F}[\{\Delta_{\vq}\}]=\mathcal{F}_0 +\sum_{\vq}\frac{\rho(E_F)}{2\lambda}\verts{\Delta_{\vq}}^2
+\sum_{l=2}^{\infty}U_l[\{\Delta_{\vq}\}],
\end{eqnarray}
where the set $\{\Delta_{\vq}\}$ contains, in principle, all harmonics of the fundamental ordering vectors,
although in practice we restrict ourselves to fourth order terms for the continuous phase transitions. 
The functionals,
$U_\ell$, are defined as
\begin{equation}
U_\ell=\frac{(-1)^{\ell+1}}{\ell}\frac{1}{\beta}\int_{0}^{\beta} \textrm{d}\tau_1\cdots \int_{0}^{\beta}\textrm{d}\tau_l \langle T_\tau V(\tau_1)\cdots V(\tau_\ell)\rangle.
\end{equation}
The expansion in terms of $U_\ell$ contains only 
fully connected, 1-loop diagrams, where
the finite-temperature Green functions for the electrons are 
\begin{eqnarray}
\mathcal{G}_0^{\eta}(\vk,\omega_n)=\frac{1}{i\omega_n-\xi_{\vk\eta}},
\end{eqnarray}
with $\omega_n = (2 n + 1)\pi/\beta$, and the order parameters, $\Delta_{\bf{Q}}$, act as
external, single-particle potentials (see, for example, Fig.~\ref{FIG-chi}).

To fourth order, the Landau free energy in terms of order parameters associated with $\vQ_1$, $\vQ_2$ and their harmonics is 
\begin{widetext}
\begin{eqnarray}\label{Landau energy}
&&\mathcal{F}=\mathcal{F}_0+\sum_{j=1}^{2} \Big\{  r_j (|\Delta_{\vQ_j}|^2+|\Delta_{\bar{\vQ}_j}|^2)
+\bar{r}_j(|\Delta_{2\vQ_j}|^2+|\Delta_{2\bar{\vQ}_j}|^2)
+\tilde{r}_j(|\Delta_{\vQ_j+\bar{\vQ}_j}|^2+|\Delta_{\vQ_j-\bar{\vQ}_j}|^2)
+ d_j (|\Delta_{\vQ_j}|^4+|\Delta_{\bar{\vQ}_j}|^4)\nonumber\\
&&~~~+ g_j |\Delta_{\vQ_1}|^2|\Delta_{\vQ_2}|^2
+ [ b_j (\Delta_{\vQ_j}^2\Delta_{-2\vQ_j} +\Delta_{\bar{\vQ}_j}^2\Delta_{-2\bar{\vQ}_2} )
+          c_j (\Delta_{\vQ_j}\Delta_{\bar{\vQ}_j}\Delta_{-\vQ_j-\bar{\vQ}_j} +\Delta_{\vQ_j}\Delta_{-\bar{\vQ}_j}\Delta_{\bar{\vQ}_j-\vQ_j})+\textrm{c.c.}] \Big\},
\end{eqnarray}
where the coefficients are given by the integrals
\begin{eqnarray}
r_j&=&\rho(E_F)/\lambda-\chi(\vQ_j;T) \\
\bar{r}_j&=&\rho(E_F)/\lambda-\chi(2\vQ_j;T) \\
\tilde{r}_j&=&\rho(E_F)/\lambda-\chi(\vQ_j+\bar{\vQ}_j;T) \\
b_j & = &  \int^{\,\prime} \mathcal{G}_0^{\eta}(0)\,\mathcal{G}_0^{\eta}(\vQ_j)\,\mathcal{G}_0^{\eta}(2\vQ_j) \\
c_j & = & \int^{\,\prime} \Big\{\mathcal{G}_0^{\eta}(0)\,\mathcal{G}_0^{\eta}(\vQ_j)\,
\mathcal{G}_0^{\eta}(\vQ_j+\bar{\vQ}_j)+\mathcal{G}_0^{\eta}(0)\,
\mathcal{G}_0^{\eta}(\bar{\vQ}_j)\,\mathcal{G}_0^{\eta}(\vQ_j+\bar{\vQ}_j)\Big\}  \\
d_j & = & \int^{\,\prime} \Big\{\mathcal{G}_0^{\eta}(0) \mathcal{G}_0^{\eta}(\vQ_j)^2 \mathcal{G}_0^{\eta}(2\vQ_j) + \frac{1}{2}\mathcal{G}_0^{\eta}(0)^2 \mathcal{G}_0^{\eta}(\vQ_j)^2\Big \} \\
g_j & = & \int^{\,\prime} \Big\{\mathcal{G}_0^{\eta}(0)^2 \mathcal{G}_0^{\eta}(\vQ_j) \mathcal{G}_0^{\eta}(\bar{\vQ}_j)
+\mathcal{G}_0^{\eta}(0)^2\mathcal{G}_0^{\eta}(\vQ_j)\mathcal{G}_0^{\eta}(-\bar{\vQ}_j)
+\mathcal{G}_0^{\eta}(0)\mathcal{G}_0^{\eta}(\vQ_j)^2\mathcal{G}_0^{\eta}(\vQ_j+\bar{\vQ}_j) \nonumber \\
&&\ \quad  + \mathcal{G}_0^{\eta}(0)\mathcal{G}_0^{\eta}(\vQ_j)^2\mathcal{G}_0^{\eta}(\vQ_j-\bar{\vQ}_j) 
+ 2\mathcal{G}_0^{\eta}(0)\mathcal{G}_0^{\eta}(\vQ_j)\mathcal{G}_0^{\eta}(\bar{\vQ}_j)
\mathcal{G}_0^{\eta}(\vQ_j+\bar{\vQ}_j)\Big\},
\end{eqnarray}
\end{widetext}
and we have adopted the compactified notation:
\begin{eqnarray}
\mathcal{G}_0^{\eta}(\vq) \equiv \mathcal{G}_0^{\eta}(\vk+\vq,i\omega_n) & \\
&\nonumber\\
\int^{\, \prime} \equiv  2\sum_{\eta} \frac{1}{\beta}\sum_{i\omega_n} \int\frac{d\vk}{(2\pi)^2}.
\end{eqnarray}

As the susceptibility towards ordering is always largest for the fundamental vectors, we integrate out
the higher harmonics, 
which allows us to write the free energy
in a form that transparently displays the $C_4$ symmetry of the problem
\begin{eqnarray}
&&\mathcal{F}=\sum_j\big[r_j(|\Delta_{\vQ_j}|^2+|\Delta_{\bar{\vQ}_j}|^2)\\
&&~~~+\frac{u_j}{4}(|\Delta_{\vQ_j}|^2+|\Delta_{\bar{\vQ}_j}|^2)^2  
+\gamma_j|\Delta_{\vQ_j}|^2|\Delta_{\bar{\vQ}_2}|^2\big].\nonumber\label{eq:final-F}
\end{eqnarray}
The effect of higher harmonics is to renormalize the coefficients of the quartic terms, which become
\begin{eqnarray}
&&u_j=4\left(d_j-|b_j|^2/\bar{r}_j\right)  \nonumber\\
&&\gamma_j=g_j-|c_j|^2/\tilde{r}_j-u_j/2.
\label{eq:u-gamma}
\end{eqnarray}


\begin{thebibliography}{33}
\bibitem[{foo({\natexlab{b}})}]{footnote2}
\bibinfo{note}{There is a small orthorhombicity in the crystal structure, which
  therefore acts as a weak symmetry breaking field. Thus, rather than domains
  of different orientation stripes, which might otherwise be expected, stripe
  order with a single ordering vector ($ \vQ_2$, not ${\bar {\vQ}_2}$) is
  stabilized.}
\bibitem[{foo({\natexlab{c}})}]{footnote3}
\bibinfo{note}{The crystal unit cell for $R$Te$_3$ is doubled from the simple
  one of the square lattice.\cite{Brouet04} It is also rotated by 45$^{\circ}$.
  If we denote the lattice spacing between nearest Te atoms by $a$, which we
  set to be 1 throughout this work, the reciprocal lattice unit $a^\ast$ in
  Ref. \onlinecite{Brouet04} is actually equal to $2\pi/\sqrt{2}$. Then
  $|\vQ_{\textrm{CDW}}|\approx 5a^\ast/7=5\pi\sqrt{2}/7$. Rotating back by
  $45^\circ$ gives, in the simple unit cell scheme used here,
  $\vQ_{\textrm{CDW}}\approx (5\pi/7,5\pi/7)$.}
\end{thebibliography}
\end{document}